# Effect of rotational-state-dependent molecular alignment on the optical dipole force


Lee Yeong Kim[1], Ju Hyeon Lee[2], Hye Ah Kim[2], Sang Kyu Kwak[3], Bretislav Friedrich[4], Bum Suk Zhao (조범석)[1,2,*]

[1]Department of Physics, Ulsan National Institute of Science and Technology, Ulsan 44919, Korea

[2]Department of Chemistry, Ulsan National Institute of Science and Technology, Ulsan 44919, Korea

[3]School of Energy and Chemical Engineering, Ulsan National Institute of Science and Technology, Ulsan 44919, Korea

[4]Fritz-Haber-Institut der Max-Planck-Gesellschaft, 14195 Berlin, Germany




## Abstract


The properties of molecule-optical elements such as lenses or prisms based on the interaction of molecules with optical fields depend in a crucial way on the molecular quantum state and its alignment created by the optical field. However, in previous experimental studies, the effects of state-dependent alignment have never been included in estimates of the optical dipole force acting on the molecules while previous theoretical investigations took the state-dependent molecular alignment into account only implicitly. Herein, we consider the effects of molecular alignment explicitly and, to this end, introduce an effective polarizability which takes proper account of molecular alignment and is directly related to the alignment-dependent optical dipole force. We illustrate the significance of including molecular alignment in the optical dipole force by a trajectory study that compares previously used approximations with the present approach. The trajectory simulations were




carried out for an ensemble of linear molecules subject to either propagating or standing-wave optical fields for a range of temperatures and laser intensities. The results demonstrate that the alignment-dependent effective polarizability can serve to provide correct estimates of the optical dipole force, on which a state-selection method applicable to nonpolar molecules could be based. We note that an analogous analysis of the forces acting on polar molecules subject to an inhomogeneous static electric field reveals a similarly strong dependence on molecular orientation.


*Corresponding author

zhao@unist.ac.kr




## 1. Introduction

Current laser technology has made it possible to generate spatially and temporally well defined optical fields – whether propagating or standing-wave – that can be used to manipulate molecular motion and create molecule-optical elements, such as lenses [1-3] and prisms [4,5], as well as to decelerate molecules [6-11]. Manipulating the translation of other than spherical molecules by optical fields entails manipulating their rotation first. The field hybridizes the rotational states of the molecules and thereby creates *directional states* in which their induced dipole moments are *aligned* with respect to the polarization vector of the field. Only in such directional states are the molecular body-fixed dipole (or higher) moments accessible in the laboratory frame and can be acted upon by space-fixed fields in order to achieve efficient manipulation of their translation. Whether the directional states will be created and hence strong optical dipole forces exerted by the optical field on the molecules is contingent upon the symmetry of the molecules' polarizability tensor: only for an anisotropic polarizability whose principal components are not all equal to one another will the hybridization of rotational states by the interaction with an optical field take place [12-16]. For instance, the polarizability tensor of every linear molecule is anisotropic, with the principal polarizability component along the molecular axis exceeding that perpendicular to it; this makes all linear molecules amenable to facile manipulation by an optical field.

In most of the previous experimental studies, a rotation-averaged molecular polarizability was used to analyze the experimental data [1,2,4,6-8,10], in which neither the effects of the rotational state nor of the molecular alignment on the optical dipole force were considered. Although it was demonstrated during the last decade that the optical dipole force is modified by the field-induced molecular alignment, the state dependence of the molecular polarizability was still ignored [9,11]. Not until very recently was the rotational-state-dependent molecular polarizability used to interpret the transverse dispersion of $CS_2$

molecules subject to pulsed optical standing waves, although the molecular alignment effect was neglected [5].

In contrast to the above experimental studies, their theoretical counterparts made use of the state dependence of molecular alignment, since the degree of molecular alignment intrinsically depends on the rotational state of the molecule [12-14,16]. Translational motion of molecules subject to propagating laser fields was traced with quantum-mechanical [17], hybrid quantum-classical [17-19] and classical [20] trajectory methods, in which the state-dependent molecular alignment was included in the Hamiltonian of the system under study. However, the trajectories were calculated without considering the relation between the molecular alignment and the optical dipole force. Therefore, in these theoretical approaches, the effect of the alignment on the optical dipole force was not explicitly included. Furthermore, these studies focused on the low-intensity [20,21] or the high-intensity limits [17-20], wherein the molecules are hardly aligned or the degree of their alignment approaches its maximum, respectively. However, in particular when molecules travel through an optical standing wave, some of them probe the wave's whole intensity range -- from zero to the peak value. Therefore, the intermediate intensity range, in which the molecules are aligned partially, is key for understanding the relation between the molecular alignment and the optical dipole force. The intensity range varies for different initial rotational states of the molecules and hence the alignment for which they intrinsically allow. For instance, at a certain laser intensity, the high-field limit is reached for molecules in the rotational ground state, whereas molecules occupying rotationally excited states may be hardly affected by the same laser field.

The rotational-state-dependent molecular alignment is a precondition for exerting the rotational-state-dependent optical dipole force, which, in turn, is the basis for selecting nonpolar molecules that occupy a specific quantum state. There has been continuous interest



in the separation and discrimination of molecules by using nonresonant laser beams [4,5,9,17,21-25], which includes isotope and spin isomer separation. New techniques based on the nonresonant optical dipole force would complement the methods for state selection of polar molecules via inhomogeneous static electric fields [26-31]. Therefore, the development of new tools for separating nonpolar molecules requires the proper evaluation and optimization of the optical rotational-state-dependent dipole force.

Based on our analysis presented herein, we introduce a new kind of effective polarizability that explicitly connects the state-dependent molecular alignment with the optical dipole force. The state-dependent variation of such an effective polarizability is directly reflected in the optical dipole force. The calculated optical dipole force is then utilized to investigate the deflection and dispersion of $CS_2$ ($X^1\Sigma$) molecules subject to either a propagating laser field or an optical standing wave as an example. The intensity of interest in this study includes the intermediate range for each rotational state as well as the low- and high-intensity ranges. We demonstrate that the new effective polarizability that includes the effect of the rotational-state-dependent alignment can be used as a general guide in assessing the optical dipole force. Therefore, our study provides an approach toward the optimization of the state-dependent optical dipole force which is a prerequisite for the quantum state selection of nonpolar molecules. We also examine the counterpart of the new effective polarizability that arises in the context of the dipole force exerted on polar molecules by an inhomogeneous static electric field.

## 2. Deflection scheme

We consider two schemes for the transverse manipulation of molecules. The first one follows the molecular deflection experiment of Refs. [1,2] in which molecules are deflected by a single laser beam propagating perpendicularly to the molecular beam, as shown in Fig.



1(a). The second scheme pertains to the experiment by Sun *et al.* [5], who studied the transverse dispersion of molecules brought about by pulsed optical standing waves, as illustrated in Fig. 1(b). In Fig. 1, "IR1" and "IR2" denote pulsed, linearly polarized infrared (IR) laser beams. In the second scheme, the requisite pulsed standing wave is generated by two counter-propagating beams IR1 and IR2, cf. Fig. 1(b). In what follows, we consider a supersonic beam of $CS_2$ molecules whose rotational temperature $T$ is assumed to be 1 or 35 K. The former temperature can be achieved by state-of-the-art molecular beam sources [32] and the latter was used in previous experimental studies [5,9,11]. The molecular beam crosses the IR beam or the optical standing wave at right angles. We choose the infrared laser beam (IR1), the molecular beam, and the laser polarization directions to be along the $x$, $z$, and $y$ axis, respectively. The coordinate origin is at the focal point of IR1 and IR2 and the time origin ($t$=0) is given by the maximum intensity of the pulsed laser beam. The respective intensities of the propagating and the standing waves can be written as follows:

$$I(\mathbf{r},t) = I_0 \exp\left[\frac{-2(y^2+z^2)}{\omega_0^2}\right] \exp\left(-4\ln(2)\frac{t^2}{\tau^2}\right) \tag{1a}$$

and

$$I(\mathbf{r},t) = 4I_0 \exp\left[\frac{-2(y^2+z^2)}{\omega_0^2}\right] \exp\left(-4\ln(2)\frac{t^2}{\tau^2}\right)\cos^2\left(\frac{2\pi}{\lambda}x\right). \tag{1b}$$

Here, $I_0$, $\omega_0$, $\tau$, and $\lambda$ are the peak intensity, waist radius ($e^{-2}$ radius), pulse duration (full width at half maximum, FWHM), and wavelength of IR1 and IR2, respectively. We choose $\omega_0 = 23.5$ $\mu$m, $\tau = 10$ ns, and $\lambda = 1064$ nm for our study. Since the rotational period $B/\hbar = 49$ ps for $CS_2$ (whose rotational constant $B$ is 0.109 cm$^{-1}$) is much smaller than the pulse duration $\tau$, the rotational motion of the molecule in the laser field is adiabatic [33].



Below, we trace the trajectories of $10^6$ molecules starting at their initial positions ($x_0$, $y_0$, $z_0$) with initial velocities ($v_{0x}$, $v_{0y}$, $v_{0z}$) at $t = -30$ ns, whose passage through one of the two laser fields results in a transverse velocity distribution $g(v_x, v_y)$ at $t = 30$ ns.

## 3. Theory

The interaction potential between a $^1\Sigma$ molecule (such as the linear ground-state $CS_2$) and a laser field of intensity $I$ is

$$U = -\frac{1}{2}(\alpha_\parallel \cos^2\theta + \alpha_\perp \sin^2\theta)Z_0 I = -\frac{1}{2}Z_0 I[(\alpha_\parallel - \alpha_\perp)\cos^2\theta + \alpha_\perp] \qquad (2)$$

where $\alpha_\parallel$ and $\alpha_\perp$ are the polarizability components parallel and perpendicular to the molecular axis, $\theta$ the polar angle between the molecular axis and the laser polarization axis (i.e., the $y$ axis), and $Z_0$ the vacuum impedance.

In the absence of the field ($I=0$), $U = 0$, in which case the molecule undergoes free rotation, represented by the Hamiltonian $\mathbf{H_0} = B\mathbf{J^2}$, with $\mathbf{J^2}$ the squared angular momentum operator and $B$ the rotational constant. The eigenfucntions of $\mathbf{J^2}$ are the spherical harmonics $Y_{j,M} = |j,M>$, pertaining to eigenvalues $j(j+1)$, with $j$ the rotational angular momentum quantum number and $M$ the projection of the angular momentum on the laser polarization axis; for $^{12}C^{32}S_2$, $j$ is restricted to even integers because of the zero nuclear spin of $^{32}S$ [34]. Within the laser field ($I>0$), the free-rotor states of the molecule get hybridized (coupled) by the $\cos^2\theta$ operator and the molecular axis aligned as a result. Under the adiabatic alignment condition, i.e., when the molecule is subject to a laser field that varies on a time scale longer than the rotational period, the wave function of the aligned molecules is obtained by solving the time-independent Schrödinger equation with Hamiltonian $\mathbf{H} = \mathbf{H_0} + U$. The corresponding solutions are superpositions of the field-free rotational states, $\Psi_{J,M}(I) = \sum_j C_j^{J,M}(I)\,|\,jM> $ [16]. $\Psi_{J,M}$ adiabatically correlates with the field-free state



$|j,M\rangle$, i.e. $\Psi_{J,M}$ evolves from $\Psi_{J,M}(I=0) = |j,M\rangle$ to $\Psi_{J,M}(I)$ when the field is turned on slowly-enough to fulfill the adiabatic condition. The expectation value of $\cos^2\theta$, $\langle\Psi_{J,M}|\cos^2\theta|\Psi_{J,M}\rangle$, characterizes the degree of the molecular alignment achieved and is denoted by $\langle\cos^2\theta\rangle_{J,M} = \langle\cos^2\theta\rangle_{J,M}(I)$, termed the alignment cosine. The alignment cosine is a rotational-state-dependent function of the laser intensity. Since the time variation of the laser field that affects molecular translation exceeds the rotational period by several orders of magnitude [17], the translational motion is governed by the following approximate potential:

$$U_{J,M}(\mathbf{r},t) = -\frac{1}{2}\alpha_{J,M}^{U}[I(\mathbf{r},t)]I(\mathbf{r},t)Z_0\,. \qquad (3)$$

Here,

$$\alpha_{J,M}^{U}(I) = (\alpha_{\parallel} - \alpha_{\perp})\langle\cos^2\theta\rangle_{J,M} + \alpha_{\perp} \qquad (4)$$

which is the polarizability component along the space-fixed laser polarization axis [18], and was termed the effective polarizability in previous experimental works [9,11]. The interaction potential, Eq. (3), and the space-fixed polarizability component, Eq. (4), depend on the degree of molecular alignment $\langle\cos^2\theta\rangle_{J,M}$ and so does the resulting dipole force $(-\nabla U_{J,M}(\mathbf{r},t))$:

$$\begin{aligned}\mathbf{F}_{J,M}(\mathbf{r},t) &= \frac{1}{2}Z_0\nabla\left\{\alpha_{J,M}^{U}[I(\mathbf{r},t)]I(\mathbf{r},t)\right\}\\ &= \frac{1}{2}Z_0\nabla I(\mathbf{r},t)\left\{\alpha_{J,M}^{U}[I(\mathbf{r},t)] + I(\mathbf{r},t)\frac{d\alpha_{J,M}^{U}[I(\mathbf{r},t)]}{dI(\mathbf{r},t)}\right\}\end{aligned}\cdot \qquad (5)$$

The two terms in the curly brackets of Eq. (5) can be considered to be components of an effective polarizability,

$$\alpha_{J,M}^{\mathbf{F}}(I) = \alpha_{J,M}^{U}(I) + \beta_{J,M}(I)\,, \qquad (6)$$

with $\alpha_{J,M}^{U}(I)$ given by Eq. (4) and

$$\beta_{J,M}(I) = I\frac{d\alpha_{J,M}^{U}(I)}{dI}\,. \qquad (7)$$



We note that the resulting dipole force $\mathbf{F}_{J,M}$ varies proportionately to the product of $\alpha_{J,M}^{\mathbf{F}}(I)$ and $\nabla I$, as the interaction potential $U_{J,M}$ is proportional to the product of $\alpha_{J,M}^{U}(I)$ and $I$. The two additional superscripts, $U$ and $\mathbf{F}$, of the two effective polarizabilities remind us of their relations to $U_{J,M}$ and $\mathbf{F}_{J,M}$, respectively. Below, we refer to the two effective polarizabilities $\alpha_{J,M}^{U}(I)$ and $\alpha_{J,M}^{\mathbf{F}}(I)$ as the $U$- and $\mathbf{F}$-effective polarizabilities, respectively.

## 4. Results and Discussion

### A. Effective polarizabilities

As shown by Eq. (5), molecular alignment modifies the optical dipole force for a state $\Psi_{J,M}$ in two ways: Firstly, the $U$-effective polarizability $\alpha_{J,M}^{U}(I)$ for $\Psi_{J,M}(I)$ varies with the laser intensity $I$. Secondly, the molecular alignment adds a new term proportional to $\beta_{J,M}(I)$, which has the same units as the polarizability ($Cm^2/V$ in SI). Therefore, the $\mathbf{F}$-effective polarizability $\alpha_{J,M}^{\mathbf{F}}(I)$, which is the sum of these two terms, encapsulates the full effect of molecular alignment on the optical dipole force. We plot $\alpha_{J,M}^{U}(I)$, $\beta_{J,M}(I)$, and $\alpha_{J,M}^{\mathbf{F}}(I)$ as a function of the laser intensity $I$ in the three columns of Fig. 2. The three rows (from top to bottom) correspond to the conditions $J \leq 4$, $J = 10$, and $J = 20$, respectively. Considering rotational state distributions for $T = 1$ and 35 K, we plotted states that are relevant to the maximum and half maximum of the distribution. For $T = 35$ K (1 K), the population of the rotational energy levels of $CS_2$ has its maximum at $J = 10$ ($J = 2$), and is close to half of the maximum at $J = 2$ and 20 ($J = 0$ and 4). For comparison, we also plot state-averaged $<\alpha_{J,M}^{U}(I)>$, $<\beta_{J,M}(I)>$, and $<\alpha_{J,M}^{\mathbf{F}}(I)>$ values for $T = 1$ (black solid line) and 35 K (red solid line) in the three graphs of the second row.



In order to characterize the variation of $\alpha_{J,M}^U(I)$, we chose three intensities, namely $I^{(+)}$, $I^{(\min)}$, and $I^{(-)}$, which are associated with the largest rising inflection point, the local minimum, and the falling inflection point close to its local minimum of the $\alpha_{J,M}^U(I)$ vs. $I$ curves, respectively. The vertical lines in Figs. 2(b) and (f) illustrate these intensities for $\Psi_{10,8}$. At this stage, $I^{(+)}$ and $I^{(-)}$ can separate the intermediate intensity range from the high and low ones, respectively. Generally, when $I > I^{(+)}$, $\alpha_{J,M}^U(I)$ starts to converge to the high-field limit:

$$\alpha_{J,M}^U(I) = \Delta\alpha\left(1 - \frac{J_i + 1}{\sqrt{\Delta\alpha Z_0 \pi / 2B}}I^{-1/2}\right) + \alpha_\perp \qquad (8a)$$

for $(J - |M|)$ even and

$$\alpha_{J,M}^U(I) = \Delta\alpha\left(1 - \frac{J_i}{\sqrt{\Delta\alpha Z_0 \pi / 2B}}I^{-1/2}\right) + \alpha_\perp \qquad (8b)$$

for $(J - |M|)$ odd [12,13]. We use respectively solid and dashed lines to distinguish between the even and the odd $(J - |M|)$ in Fig. 2. The local minimum exists for most rotational states except for those with a high ratio of $|M|$ to $J$. For example, there is no $\alpha_{J,M}^U(I)$ minimum for $|M| = J$ and $J - 1$ as long as $J > 2$. Therefore, $I^{(\min)}$ and $I^{(-)}$ are not defined for those states.

In Fig. 3, these three intensities are plotted together against $J$ for (a) $|M| = 0$, $J$ and (b) 1, $J - 1$. Since only even $J$ values occur for $^{12}C^{32}S_2$ molecules [34], $J - |M|$ is even (odd) when $|M| = 0$ or $J$ (1 or $J - 1$). $I^{(+)}$ increases with $J$, which explains the fact that a stronger laser intensity is necessary for high-$J$ rotational states to reach the high-field limit because of their large rotational energy. Therefore, as shown in Figs. 2(a) and (b), the $\alpha_{J,M}^U(I)$ values for the states of $J \leq 10$ converge to their corresponding asymptotic values given by Eq. (8) at $I = 10^{12}$ W/cm$^2$, whereas the $\alpha_{20,M}^U(I)$ value at the same intensity is quite different from the corresponding high-field value. This is related to the fact that the intensity of $10^{12}$ W/cm$^2$ is smaller than $I^{(+)}$ for the rotational states of $J = 20$. $I^{(\min)}$ and $I^{(-)}$ also increase as $J$ increases.



Thus, $I = 10^{12}$ W/cm$^2$ corresponds to the high-field limit for the states of $J \leq 10$, but to the intermediate range for the state with $J = 20$.

Figures 3(c) and (d) show the variations of $I^{(+)}$, $I^{(\min)}$, and $I^{(-)}$ as a function of $|M|$ for $J = 10$ and 16. Here, the states for even and odd $(J - |M|)$ values are plotted in Figs. 3(c) and (d), respectively. All three characteristic intensities decrease as $|M|$ increases. This tendency of $I^{(+)}$ is counterintuitive, since the rotational states of $|M| \approx 0$, whose rotational plane includes the laser polarization axis, are expected to be aligned by the lower laser intensity. We tentatively attribute this behavior to the inclination in $I^{(\min)}$, which is closely correlated to $I^{(+)}$.

This behavior comes about as follows: As the laser intensity $I$ increases, the contribution of $|j,M>$ with larger $|J - j|$ to $\Psi_{J,M}(I)$ becomes more important. Since the degree of alignment of a field-free state $|j,M>$, $<j,M|\cos^2\theta|j,M>$, becomes larger as $j$ increases (for a given $M$), the contribution of $|j,M>$ states with $j > J$ ($j < J$) increases (decreases) $<\cos^2\theta>_{J,M}$ and thus $\alpha_{J,M}^U(I)$. Therefore, $<\cos^2\theta>_{J,J}$ and $\alpha_{J,J}^U(I)$ always increase with the laser intensity, since there is no $|j,M=J>$ state with $j < J$. For $\Psi_{J,M}(I)$ states with same $J$ and different $|M|$ values, at small $|M|$ more $|j,M>$ states with $j < J$ contribute to $\Psi_{J,M}(I)$. Since the hybridization of a given $\Psi_{J,M}(I)$ state with $|j,M>$ states is easier for $j < J$ than for $j > J$ (due to the decreasing and increasing energy differences), the increase of the laser intensity reduces $\alpha_{J,M}^U(I)$ in low-intensity range. When the intensity increases further, the additional contribution from $|j,M>$ states with larger $j$ balances out the low-$j$ state contribution, which results in a minimum of $\alpha_{J,M}^U(I)$ at $I^{(\min)}$. Therefore, $I^{(\min)}$ decreases with $|M|$ increasing, and so does $I^{(+)}$.

In the presence of molecular alignment, the second term of Eq. (6), $\beta_{J,M}(I)$, is essential for estimating the optical dipole force of Eq. (5) ($\beta_{J,M}(I) = 0$ when the alignment is ignored). Therefore, we plot it separately in the middle column of Fig. 2. Since $\beta_{J,M}(I)$ is the product of $I$ and $d\alpha_{J,M}^U(I)/dI$, the variation of $\alpha_{J,M}^U(I)$ with respect to $I$ is magnified by the laser



intensity $I$. Thus, the fluctuation amplitude of $\beta_{J,M}(I)$ increases with $J$, although that of $d\alpha_{J,M}^{U}(I)/dI$ decreases with $J$. For these reasons, it could be misleading to consider only $d\alpha_{J,M}^{U}(I)/dI$ without the multiplication factor $I$ for evaluating the optical dipole force acting on the molecules. The three characteristic intensities $I^{(+)}$, $I^{(\min)}$, and $I^{(-)}$ are also helpful for understanding the behavior of $\beta_{J,M}(I)$. Since $\beta_{J,M}(I^{(\min)}) = 0$ according to Eq. (7), $\beta_{J,M}(I)$ is usually negative (positive) when $I < I^{(\min)}$ ( $I > I^{(\min)}$ ). Although $\beta(I^{(\pm)})_{J,M}$ is not exactly equal to the two main extrema of $\beta_{J,M}(I)$, they closely follow $I^{(-)}$ and $I^{(+)}$. Furthermore, when $I^{(\pm)}$ increases as $J$ increases with $|M|$ fixed (Figs. 3(a) and (b)) or as $|M|$ decreases with $J$ fixed (Figs. 3(c) and (d)), the difference between the maximum and the minimum of $\beta_{J,M}(I)$ becomes increasingly amplified, which is well illustrated in Figs. 2(d)–(f). For example, the difference is about $5 \times 10^{-40}$ Cm$^2$/V for $\beta_{4,0}(I)$, as shown in Fig. 2(d), which increases up to $15 \times 10^{-40}$ Cm$^2$/V for $\beta_{20,0}(I)$, as shown in Fig. 2(f). The variations of $\beta_{10,M}(I)$ for even and odd $|M|$ also exemplify this aspect very well. In each group, the difference increases as $|M|$ decreases. The properties of $\beta_{J,M}(I)$ at $I >> I^{(+)}$ are intriguing. In this high-field limit, we obtain:

$$\beta_{J,M}(I) = \frac{\Delta\alpha}{2} \frac{J+1}{\sqrt{\Delta\alpha Z_0 \pi / 2B}} I^{-1/2} \qquad (9a)$$

for $(J - |M|)$ even and

$$\beta_{J,M}(I) = \frac{\Delta\alpha}{2} \frac{J}{\sqrt{\Delta\alpha Z_0 \pi / 2B}} I^{-1/2} \qquad (9b)$$

for $(J - |M|)$ odd. The coefficient of $I^{-1/2}$ in this equation differs by a factor of $-1/2$ from the coefficient in Eq. (8). Thus, $\beta_{J,J}(I)$ is larger than $\beta_{J,J-1}(I)$, while $\alpha_{J,J}^{U}(I)$ is smaller than $\alpha_{J,J-1}^{U}(I)$. In addition, the separation between $\beta_{J,J}(I)$ and $\beta_{J,J-1}(I)$ is half as large as the separation between $\alpha_{J,J}^{U}(I)$ and $\alpha_{J,J-1}^{U}(I)$.



Considering the variations of $\alpha_{J,M}^{U}(I)$ and $\beta_{J,M}(I)$, we can explain the behavior of $\alpha_{J,M}^{F}(I)$, which is directly correlated with that of the optical dipole force. In the high-field limit, since the signs of $\beta_{J,J}(I) - \beta_{J,J-1}(I)$ and $\alpha_{J,J}^{U}(I) - \alpha_{J,J-1}^{U}(I)$ are opposite, $\alpha_{J,M}^{F}(I)$ is more tightly spaced than $\alpha_{J,M}^{U}(I)$. Because of the wild variation of $\beta_{J,M}(I)$, for some states (such as $\Psi_{10,0}(I)$, $\alpha_{J,M}^{F}(I)$ at an intermediate intensity of $I = 0.4 \times 10^{12}$ W/cm$^2$ can be larger than $\alpha_{J,M}^{F}(I)$ at the high-field limit of $I = 10^{12}$ W/cm$^{-2}$, as shown in Fig. 2(h). Additionally, the states of the same $J$ and different $|M|$ values have the broad distribution of $\alpha_{J,M}^{F}(I)$ at the intermediate intensity range for $|M| = 0$. In particular, $\alpha_{J,M}^{F}(I)$ for different $|M|$ values ranges from $2.5 \times 10^{-40}$ to $16 \times 10^{-40}$ Cm$^2$/V at $I = 10^{12}$ W/cm$^2$.

The approach used to develop the F-effective polarizability can also be applied to the interaction between a dipole moment and an inhomogeneous static electric field $E(\mathbf{r})$, whose potential is given by:

$$U_{S}(\mathbf{r}) = -\mu_{\text{eff}}[E(\mathbf{r})]E(\mathbf{r}). \tag{10}$$

Here, $\mu_{\text{eff}}$ is the space-fixed electric dipole moment. The interaction results in a dipole force given by:

$$\begin{aligned}\mathbf{F}_{S}(\mathbf{r}) = -\nabla U_{S}(\mathbf{r}) &= \mu_{\text{eff}}[E(\mathbf{r})]\nabla E(\mathbf{r}) + E(\mathbf{r})\nabla \mu_{\text{eff}}[E(\mathbf{r})] \\ &= \nabla E(\mathbf{r})\left\{\mu_{\text{eff}}[E(\mathbf{r})] + \frac{d\mu_{\text{eff}}}{dE}E(\mathbf{r})\right\}\end{aligned}. \tag{11}$$

The explicit formula for the dipole force has been approximated by $\mu_{\text{eff}}\nabla E(\mathbf{r})$ [28,30,31] by neglecting the second term $(d\mu_{\text{eff}}/dE)E$, which is a counterpart of $\beta_{J,M}(I)$ in the optical dipole force. However, the magnitudes of the first and second terms are the same when $\mu_{\text{eff}}$ is a linear function of $E$. The absolute ground states of para and ortho water [35] can be approximated using this condition. Furthermore, the second term which is, by the Hellmann-Feyman theorem [15,36], proportional to the orientation cosine, can be an order of magnitude



larger than the first term for the iodobenzene molecules in the rotational states of $J_{K_a K_c} M =$ $3_{03}2$ and $4_{13}3$ whose $\mu_{\text{eff}}$ values increase sharply at electric field strengths of about 30 kV/cm and 55 kV/cm, respectively. (See Fig. 6 in Ref. [29].)

The explicit evaluation of both the induced-dipole force due to an optical field, Eq. (5), and the permanent-dipole force due to an inhomogeneous electrostatic field, Eq. (11), offers advantages in analyzing and tailoring the respective contributions from the first and second terms of Eqs. (5) and (11). For an accurate eigenenergy surface spanned by the spatial coordinates and pertaining to a given quantum state, the force is simply the (negative) gradient of the surface for both the permanent and induced dipole interactions. However, in such implicit approach [17-19,29]) the insight about the relative contribution of the two terms to the dipole force is lost. In contrast, the explicit approach, presented here, reveals the relative roles of the two terms in the dipole force quite clearly. For example, although the optical induced-dipole force for $\Psi_{10,8}$ at intensities of $I = 0.3 \times 10^{12}$ and $1.0 \times 10^{12}$ W/cm$^2$ is the same, the second, alignment-dependent term contributions are 33% and 13%, respectively. As for the permanent dipole force, the case of iodobenzene is a good example: an implicit analysis may overlook that the contribution from the second, orientation-dependent term exceeds that from the first by an order of magnitude. By tuning the fields in the vicinity of abrupt changes of the dipole force, we can find the loci of the avoided-intersections of eigenenergies that are the cause of the abrupt changes.

## B. Interaction potential and optical dipole force

With the two effective polarizabilities $\alpha_{J,M}^U(I)$ and $\alpha_{J,M}^{\mathbf{F}}(I)$, we can calculate the molecular interaction potential and the optical dipole force, respectively, for each of the two schemes shown in Fig. 1. The molecular interaction potentials for the cases of the propagating wave field and the optical standing wave are as follows:



$$U_{J,M}(y,z,t) = -\frac{1}{2}\alpha_{J,M}^{U}[I(y,z,t)]Z_0 I_0 \exp\left[\frac{-2(y^2+z^2)}{\omega_0^2}\right]\exp\left(-4\ln(2)\frac{t^2}{\tau^2}\right) \quad (12a)$$

and

$$U_{J,M}(x,y,z,t) = -2\alpha_{J,M}^{U}[I(x,y,z,t)]Z_0 I_0 \exp\left[\frac{-2(y^2+z^2)}{\omega_0^2}\right]\exp\left(-4\ln(2)\frac{t^2}{\tau^2}\right)\cos^2\left(\frac{2\pi}{\lambda}x\right),$$

$$(12b)$$

respectively. In order to illustrate the spatial variation of the two potentials, we consider the potential $U_{J,M}(z)$ along $y = \omega_0/2$ at $t = 0$ for the former, Eq. (12a), and the potential $U_{J,M}(x)$ along the $x$ axis, at $t = 0$ for the latter, Eq. (12b), i.e.:

$$U_{J,M}(z) = -\frac{1}{2}\alpha_{J,M}^{U}[I(z)]Z_0 I_0 e^{-1/2}\exp\left(\frac{-2z^2}{\omega_0^2}\right) \quad\quad (13a)$$

and

$$U_{J,M}(x) = -2\alpha_{J,M}^{U}[I(x)]Z_0 I_0 \cos^2\left(\frac{2\pi}{\lambda}x\right), \quad\quad (13b)$$

respectively. Concentrating on the change of the transverse velocity of the molecules, we examine the following optical dipole forces:

$$F_{J,M}^{y}(z) = -\frac{\partial U_{J,M}}{\partial y}\bigg|_{y=\omega_0/2, t=0} = -\alpha_{J,M}^{F}[I(z)]Z_0\frac{1}{\omega_0}I_0 e^{-1/2}\exp\left(\frac{-2z^2}{\omega_0^2}\right) \quad (14a)$$

and

$$F_{J,M}^{x}(x) = -\frac{\partial U_{J,M}}{\partial x}\bigg|_{y=z=t=0} = -2\alpha_{J,M}^{F}[I(z)]Z_0\frac{2\pi}{\lambda}I_0\sin\left(\frac{4\pi}{\lambda}x\right). \quad\quad (14b)$$

The effect of rotational-state-dependent alignment on the interaction potential and the optical dipole force is studied in two ways: by increasing $J$ with $M = 0$ and by increasing $|M|$ for a given $J$. In Figs. 4-6, we depict the negative interaction potential and the corresponding dipole force using black and red solid curves, respectively. For comparison, we also plot the two functions calculated without considering the molecular alignment, namely, using



$\alpha_{J,M}^U(0)$ and $\alpha_{J,M}^{\mathbf{F}}(0) = \alpha_{J,M}^U(0)$ instead of $\alpha_{J,M}^U(I)$ and $\alpha_{J,M}^{\mathbf{F}}(I)$ for $U_{J,M}$ and $F_{J,M}$, respectively. The black and red dotted lines denote these results, respectively. The pair of numbers (for example, (0, 0) in the first graph) at the upper-left corner of each graph are the $J$ and $M$ values of a selected rotational state $\Psi_{J,M}$.

Figure 4 illustrates the effect of the molecular alignment induced by the propagating-wave field on $U_{J,M}(z)$ and $F_{J,M}^{y}(z)$. In the first column, we compare the potentials and the optical dipole forces for rotational states of different $J = 0, 2, 4, 6, 10$ and $20$ and the same $M = 0$. In the second and the third columns, the $M$-dependences are shown for the states of $J = 10$ and $20$, respectively. In order to remain below the ionization threshold of the $CS_2$ molecule, we choose the peak intensity $I_0$ of $10^{12}$ W/cm$^2$ in our calculation. Then, the maximum intensity along $y = \omega_0/2$ is $6.1 \times 10^{11}$ W/cm$^2$, which generally determines the peak values of $U_{J,M}(z)$ and $F_{J,M}^{y}(z)$.

The molecular alignment affects $U_{J,M}(z)$ and $F_{J,M}^{y}(z)$ via $\alpha_{J,M}^U(I)$ and $\alpha_{J,M}^{\mathbf{F}}(I)$, respectively. The series of graphs in Fig. 4 exemplifies three general aspects of the state-dependent molecular alignment effect. (A) The difference between the solid and dotted lines reflects the ratios of $\alpha_{J,M}^U(I)/\alpha_{J,M}^U(0)$ for the potential and $\alpha_{J,M}^{\mathbf{F}}(I)/\alpha_{J,M}^U(0)$ for the dipole force. This appears in those graphs in various ways. For instance, the solid lines of $U_{0,0}(z)$ and $F_{0,0}^{y}(z)$ show the largest difference from their dotted counterpart at $z = 0$ where $I = 6.1 \times 10^{11}$ W/cm$^2$, since $\alpha_{0,0}^U(I)$ and $\alpha_{0,0}^{\mathbf{F}}(I)$ increase most strongly from their field free values at the given intensity. When $\alpha_{J,M}^{\mathbf{F}}(I) = \alpha_{J,M}^U(0)$ and $\alpha_{J,M}^U(I) = \alpha_{J,M}^U(0)$, the solid and dotted lines cross each other. Furthermore, since $\alpha_{10,0}^{\mathbf{F}}(I)$ fluctuates most with respect to its field-free value in the intensity range of $0 < I < 6.1 \times 10^{11}$ W/cm$^2$, the red solid line of $F_{10,0}^{y}(z)$ oscillates most strongly with respect to the corresponding red dotted line. On the other hand,



the black solid line of $U_{10,0}(z)$ shows moderate variation from the corresponding dotted line, and the maxima of the black solid and dotted lines are almost equal, which is related to the weak variation of $\alpha_{10,0}^{U}(I)$ within the same intensity range.

(B) The optical dipole force fluctuates more widely than the interaction potential since $\alpha_{J,M}^{F}(I)$ varies more severely than $\alpha_{J,M}^{U}(I)$ because of the additional term $\beta_{J,M}(I)$, as shown in Fig. 2. This aspect can also be seen in the two graphs for $J = 10$ with $M = 1$ and 2. Moreover, for the states of $J = 20$ with $M = 16$ and 18, the size of the optical force is reduced by the molecular alignment at the peak intensity, while the negative interaction potential is slightly enhanced. (C) The variations of $\alpha_{J,M}^{U}(I)$ and $\alpha_{J,M}^{F}(I)$ occurring at high laser intensities appear very distinctly in the potential and the force, respectively, while those occurring at low intensities are hardly visible in these functions; that is because $\alpha_{J,M}^{U}(I)$ and $\alpha_{J,M}^{F}(I)$ are magnified by factors of $I$ and $\nabla I$, respectively, at a given intensity. The largest changes of $\alpha_{0,0}^{U}(I)$ and $\alpha_{0,0}^{F}(I)$ ($\alpha_{2,0}^{U}(I)$ and $\alpha_{2,0}^{F}(I)$) occur below $I = I^{(+)} = 0.4 \times 10^{10}$ W/cm$^2$ ($I^{(+)} = 3.5 \times 10^{10}$ W/cm$^2$). Thus, the shape changes of $U_{0,0}(z)$ and $F_{0,0}^{y}(z)$ ($U_{2,0}(z)$ and $F_{2,0}^{y}(z)$) are not visible in the given $y$ axis scale of the graph. On the other hand, the noticeable appearance changes in $U_{10,0}(z)$ and $F_{10,0}^{y}(z)$ can be attributed to the substantial variation of $\alpha_{10,0}^{U}(I)$ and $\alpha_{10,0}^{F}(I)$ occurring around $I = I^{(+)} = 38 \times 10^{10}$ W/cm$^2$. Additionally, the oscillation of $F_{10,M}^{y}(z)$ gets weaker, as shown in the second column of Fig. 4, when $I^{(+)}$ decreases with $|M|$ increasing within the states of the even $J - |M|$ value, as shown in Fig. 3(c).

The molecular alignment induced by the optical standing wave field also affects the interaction potential $U_{J,M}(x)$ and the optical dipole force $F_{J,M}^{x}(x)$, but there are two important differences. (i) Whereas the extremum positions of $U_{J,M}(z)$ and $F_{J,M}^{y}(z)$, which are Gaussian



functions, are identical, the maximum position of $F_{J,M}^x(x)$ is shifted by about $3\lambda/4$ from that

of $-U_{J,M}(x)$. Therefore, $F_{J,M}^x(x)$ has its maximum and minimum values at $I = 2I_0$, while the

maximum and the minimum of $-U_{J,M}(x)$ occur at $I = 4I_0$ and 0, respectively. (ii) The

maximum optical dipole force induced by the optical standing wave is orders of magnitude

larger than the one induced by the single propagating wave, when their potential depths are

the same. From Eq. (14), we can deduce that the maximum magnitude of $\nabla I$ is associated

with $1/\omega_0$ and $2\pi/\lambda$ for the propagating and the standing fields, respectively. Since $\omega_0 \approx 20$

$\mu$m and $\lambda \approx 1$ $\mu$m in our schemes, the ratio of the two maxima is about $1/100$.

Considering these two differences, we estimate the molecular alignment effect on

$U_{J,M}(x)$ and $F_{J,M}^x(x)$. In Fig. 5, we choose again $J = 0, 2, 4, 6, 10$, and 20 with $M = 0$ to study

the $J$ dependence. To investigate the $M$ dependence, we plot $U_{10,M}(x)$ and $F_{10,M}^x(x)$ for $M = 1$,

2, 9, and 10 in Fig. 6. Since the potential and the force are periodic along the $x$ axis, we plot

them for the range of one wavelength ($\lambda = 1064$ nm). The numbers on top of each column are

the value of $I_0$ used in the calculations, which are 5, 10, and 20 × $10^{10}$ W/cm$^2$.

The general aspects (namely, (A)–(C)) observed in Fig. 4 also appear in Figs. 5 and 6

together with the two differences (namely, (i) and (ii)). Concerning aspect (A) and difference

(i), $\alpha_{J,M}^U(4I_0) / \alpha_{J,M}^U(0)$ and $\alpha_{J,M}^F(2I_0) / \alpha_{J,M}^U(0)$ are responsible for the differences between

the solid and dotted lines of the potential and the optical dipole force, respectively, at their

peak positions. Therefore, if $I^{(\pm)}$ for a given state is closer to $2I_0$, then the red solid line shows

an anomalous shape at its peak position. Furthermore, aspect (B) and difference (ii) are

responsible for the distinctive step-like feature of $F_{4,0}^x(x)$ and the weak variation of $U_{4,0}(x)$ at

$I_0 = 5 \times 10^{10}$ W/cm$^2$ in Fig. 5. The combination of aspect (C) and difference (i) accounts for

the weakening of the fluctuation of $F_{J,M}^x(x)$ at $x = 0$ and $\pm\lambda/2$ where $I = 4I_0$. This is



illustrated in the red solid curves in the first column of Fig. 6. Since the standing wave intensity is $4I_0 = 20 \times 10^{10}$ W/cm$^2$ at $x = 0$ and $\pm\lambda/2$, and is similar to $I^{(-)}$ of $\alpha_{10,M}^{U}(I)$ (see Figs. 3(c) and (d)), $\alpha_{J,M}^{\mathbf{F}}(I)$ fluctuates strongly there. However, $F_{10,M}^{x}(x)$ shows very weak variation at those positions. Lastly, difference (ii) can explain the fact that while the potential depths of $U_{10,M}(x)$ in the second column of Fig. 6 are slightly lower than those of $U_{10,M}(z)$ in the second column of Fig. 4, the maximum of $|F_{10,M}^{x}(x)|$ is 50 times larger than that of $|F_{10,M}^{y}(z)|$.

The reflection of the wild variation of the new effective polarizability $\alpha_{J,M}^{\mathbf{F}}(I)$ on the optical dipole forces significantly depend on the type of the optical field. Moreover, the effect of the state-dependent molecular alignment is revealed by the difference between the solid and dotted lines. For the further analysis, the optical dipole force can be converted to an observable such as a molecular velocity [1,2,5,9].

## C. Velocity distributions

In this section, we make use of the optical dipole force that depends on the rotational-state-dependent molecular alignment to calculate the transverse velocity distribution $g(v_x, v_y)$ of molecules that have passed through the propagating wave or the optical standing wave, cf. Fig. 1. The velocity change along each axis is given by:

$$\Delta v_i = \int \frac{1}{m} F_{J,M}^{i}(x, y, z, t) dt \ \ (i = x, y, z). \tag{15}$$

Since, for the first scheme in Fig. 1, we neglect the $x$-dependence of the propagating laser intensity in Eq. (1a), we use the velocity profile $h(v_y)$ along the $v_y$ axis as a proxy for $g(v_x, v_y)$. In the second scheme – i.e. that is the molecular dispersion caused by the optical standing wave with a relatively low $I_0$ – we consider molecules passing near $y = 0$, which allows us to use the velocity profile $h(v_x)$ along the $v_x$ axis [5].



The Monte Carlo sampling method is used to select the initial velocity ($v_{0x}$, $v_{0y}$, $v_{0z}$), the initial position ($x_0$, $y_0$, $z_0$), and the initial rotational state $|J,M>$ of each individual molecule. We follow the approximation used in the previous work by Sun *et al*. [5] for the initial velocity and position distributions. A two-dimensional Gaussian function with FWHMs of $\Delta v_x^{\text{init}} = 7.2$ and $\Delta v_y^{\text{init}} = 3.4$ m/s represents the initial transverse velocity distribution. The probability function for $v_{0z}$ is a Gaussian function with the most probable velocity $v_{mp}$ of 560 m/s and a FWHM of 56 m/s. We determine $z_0$ from $v_{0z}$ according to the following equation: $z_0 = v_{mp}t_{\text{detection}} - v_{0z}t_{\text{simul}}$. Here, $t_{\text{detection}}$ and $t_{\text{simul}}$ are the detection and the total simulation time, respectively. From this initial point, the individual molecule arrives at the detection plane $z = v_{mp}t_{\text{detection}}$ at $t = t_{\text{detection}}$. $x_0$ and $y_0$ are chosen randomly from a 600 $\mu$m $\times$ 3 $\mu$m rectangle, whose $y$-center is set to $\omega_0/2$ and 0 for the first and the second schemes, respectively. The initial rotational state follows the Boltzmann distribution $e^{-Bj(j+1)/kT}/q_r$, where $k$ is Boltzmann's constant and $q_r$ the rotational partition function.

Figures 7(a) and (b) show the velocity distribution $h(v_y)$ of the deflected molecules by the propagating laser beams for $I_0 = 10^{12}$ W/cm$^2$ and $T = 35$ and 1 K, respectively. In each graph, the red solid curve shows $h(v_y)$ simulated with the alignment-included, state-dependent polarizability $\alpha_{J,M}^{\text{F}}(I)$; the blue dashed curve is obtained with the alignment-included, state-averaged polarizability $<\alpha_{J,M}^{\text{F}}(I)>$; and the black dotted curve is due to a simulation with the alignment-ignored, state-dependent polarizability $\alpha_{J,M} = \alpha_{J,M}^{\text{F}}(0)$. The initial velocity distribution is illustrated by the gray curve. All distributions are shifted toward the optical field ($\Delta v_y < 0$), reflecting the attractive character of the optical force near $y = \omega_0/2$. The distributions obtained by considering the molecular alignment (red solid and blue dotted curves) are shifted further, as expected from the polarizability enhancement. For $T = 35$ K, the width of the red solid curve is two times broader than that of the blue dotted curve,



although the weighted centers of the two distributions are identical. On the other hand, both the widths and the weighted centers of the two distributions are almost the same for $T = 1$ K. Furthermore, the weighted center of the red solid line calculated for $T = 1$ K is shifted further toward slower velocities than the one calculated for $T = 35$ K.

In order to explain these aspects, we plot the $v_y(z)$ value of various initial rotational states with $v_{0y}$, $v_{0z}$, and $y_0$ set to 0, $v_{mp}$, and $\omega_0/2$, respectively. Figures 8(a)–(c) show $v_y(z)$ for $J \leq 4$, $J = 10$, and $J = 20$, respectively. The red and black solid curves in Fig. 8(b) illustrate $v_y(z)$ values calculated using the state-averaged $<\alpha_{J,M}^{\mathrm{F}}(I)>$ values for $T = 35$ and 1 K, respectively. Generally, the final velocities of the states with same $J$ group together, while the corresponding mean velocity increases from −13.0 m/s for $J = 0$ to −7.7 m/s for $J = 20$. The mean final velocities for $J = 2$, 4, and 10 are −12.2, −11.3, and −8.8 m/s, respectively. Within each group, the final velocity clusters according to the $(J - |M|)$ value of the states, as long as the peak intensity along the molecular beam path reaches the high-field limit for $\alpha_{J,M}^{U}(I)$. However, when the high-field limit is not reached, the solid and dotted curves are mixed together, as illustrated for $J = 20$ in Fig. 8(c). The odd and even values of $(J - |M|)$ are indicated by the dashed and solid curves in the graphs, respectively. For the former case, the average final velocity of the even-valued state is about 0.5 m/s larger than that of the odd valued. Their standard deviation is almost identical for the even and odd-valued states, and increases from 0.005 m/s for $J = 0$ to 0.13 m/s for $J = 10$. Within the cluster of the same $(J - |M|)$ value, the final velocity becomes more negative as $|M|$ decreases. It is noteworthy that the effect of increasing $J$ for a fixed $|M|$ value on the final velocity is more significant than the effect obtained by changing $|M|$ while keeping $J$ and the $(J - |M|)$ value constant. A previous theoretical study [17] that did not consider the sorting of states according to whether their $(J - |M|)$ was even or odd predicted the opposite behaviour.



From the rotational distribution of CS$_2$ molecules at a given temperature, we know that the final velocities of the molecules occupying rotational states with $J \leq 20$ and $\leq 4$ contribute significantly to the $h(v_y)$ profile shown in Figs. 7(a) and (b), respectively. Therefore, the distribution of the final velocities for different $J$ values up to 20 is responsible for the broader width of the red solid curve in Fig. 7(a). On the other hand, since the molecules occupy only a few rotational states at $T = 1$ K, and the FWHM of the calculated final velocity in Fig 8(a) is roughly 2 m/s, the width of $h(v_y)$ is determined by the FWHM of the initial velocity distribution $\Delta v_y^{\text{init}} = 3.4$ m/s. Accordingly, the red solid and the blue dashed curves overlap in Fig. 7(b). Furthermore, the degree of the state-dependent alignment is larger for low-$J$ states, which results in a larger velocity change as shown in Fig. 8(a). Therefore, the red solid curve for $T = 1$ K is shifted further than the one pertaining to $T = 35$ K, at which temperature the less-aligned molecules in high-$J$ states dominate the $h(v_y)$ distribution.

Using the same format as Fig. 8, Fig. 9 shows the velocity distribution $h(v_x)$ of the molecules dispersed by the optical standing waves for $T = 35$ (upper panel) and 1 K (lower panel). Four $I_0$ values of $1.0 \times 10^{10}$, $5.0 \times 10^{10}$, $10.0 \times 10^{10}$, and $20.0 \times 10^{10}$ W/cm$^2$ are chosen for the calculation of the first, second, third, and fourth columns, respectively. For $T = 35$ K and for $I_0$ values up to $10.0 \times 10^{10}$ W/cm$^2$, the molecular alignment hardly changes the velocity distributions. In other words, the red solid and black dotted curves are very similar to each other. The main feature of $h(v_x)$, which is the averaging-out of inner rainbow-like peaks [5], is the same for the red solid and the black dotted curves. Additionally, their peak values at $x = 0$ are almost identical. These two features confirm that the approximation in the previous work, in which the molecular alignment was neglected, is appropriate [5]. The optical dipole force calculated in Figs. 5 and 6 supports this result. In particular, at $I_0 = 5.0 \times 10^{10}$ W/cm$^2$, which is the condition for Fig 9(b), only the optical dipole forces $F_{J,M}^x(x)$ for



the states of $J = 0$ and 2 calculated with $\alpha_{J,M}^{\mathrm{F}}(I)$ (red solid curves) are significantly different from those calculated with $\alpha_{J,M}^{\mathrm{F}}(0)$ (red dotted curves). The fraction of molecules in those rotational states is just 0.05 for $T = 35$ K. In addition, since the optical dipole force in Figs. 5 and 6 is the largest one in the time domain, the time-averaged optical dipole force is less affected by the molecular alignment. However, in the blue dashed curve, the rainbow-like peaks are quite distinctive, and therefore the alignment-dependent, state-averaged polarizability $<\alpha_{J,M}^{\mathrm{F}}(I)>$ is not suitable for analyzing the velocity distribution. When $I_0 = 20.0 \times 10^{10}$ W/cm$^2$, we cannot ignore the alignment effect. The red solid and black dotted curves are quite different in Fig. 9(d); more specifically, the inner rainbow-like peaks are washed out in different ways, and the peak value of the red solid curves is 30% smaller than that of the black dotted curve.

In contrast, the rainbow-like peaks are not washed out in the red solid curves for $T = 1$ K, as shown in Figs. 9(e)–(h). Furthermore, the blue dashed curves are closer to the red solid curves than the black dotted curves. The former results from the fact that the molecules occupy only a few rotational states at $T = 1$ K, and their $\alpha_{J,M}^{\mathrm{F}}(I)$ values (curves in Fig 2 (g)) are relatively similar to the $<\alpha_{J,M}^{\mathrm{F}}(I)>$ value (black solid curve in Fig. 2(h)) throughout the whole intensity range of our interest. $\alpha_{J,M}^{\mathrm{F}}(I)$ for the populated states at $T = 1$ K is significantly larger than $\alpha_{J,M}^{\mathrm{F}}(0)$ for these states even at the relatively low laser intensity, which accounts for the deviation of the red solid curves from the black dotted ones.

## 5. Conclusion

We have studied how state-dependent molecular alignment induced by two kinds of laser fields (propagating- and standing-wave) affects the molecular interaction potential, the



optical dipole force, and the resulting velocity distribution as obtained by trajectory simulations. In addition to the effective polarizability $\alpha_{J,M}^{U}(I)$ related to the potential, we introduced an effective polarizability $\alpha_{J,M}^{F}(I)$, which is explicitly related to the optical dipole force and that differs from $\alpha_{J,M}^{U}(I)$ by a rotational-state and intensity dependent term $\beta_{J,M}(I)$. As such, it provides a more direct estimate of the optical dipole force compared to the effective polarizabilities $\alpha_{J,M}^{U}(0)$ [5], $<\alpha_{J,M}^{F}(I)>$ [9,11], or $\alpha_{J,M}^{U}(I)$ and enables a more accurate and insightful analysis of the transversal motion of the molecules induced by the laser field. We also investigated the dipole force acting on a polar molecule in an inhomogeneous static electric field and found that it involves a second term similar to $\beta$. For many molecules and experimental conditions, the second term is comparable to the first term in the dipole force.

Our work therefore provides a way for estimating and tailoring the state-dependent optical dipole force more accurately, thus paving the way for developing separation techniques that can be applied to any polarizable molecules, including nonpolar ones: mixtures of nonpolar conformers, isotopes of homonuclear diatomic molecules, or of their spin isomers.



**Acknowledgement**


This work was supported by a grant from the National Research Foundation of MEST, Korea (NRF-2015R1A2A2A01005458).




**FIG. 1.** Two schemes for the transverse-manipulation of molecules. (a) A single propagating infrared (IR) laser beam and (b) a pulsed optical standing wave are used for the deflection and dispersion of molecular beams, respectively. The $CS_2$ molecular beam propagates along the $z$ axis, while the IR laser beams propagate along the $x$ axis. The coordinate origin is the focal point of the IR laser beams; the time zero is the moment when the intensity of the pulsed laser beam has its maximum value. The gray transparent beams indicate the molecular beam with the laser field off, while the pink beams indicate the molecular beam with the laser field on. The black dashed line represents initial $y$-position of the molecular beam.

**FIG. 2.** $\alpha_{J,M}^{U}(I)$, $\beta_{J,M}(I)$, and $\alpha_{J,M}^{F}(I)$ as a function of intensity for the selected rotational states. The first, second, and third rows are for $J \leq 4$, $J = 10$, and $J = 20$, respectively. For the same $J$ value, curves for different $|M|$ values are distinguished by different line brightnesses (i.e., darker curves represent larger $|M|$ values). The solid and dashed curves distinguish the states with even and odd ($J - |M|$) values, respectively.

**FIG. 3.** Three characteristic intensities $I^{(+)}$ (square and solid line), $I^{(min)}$ (circle and dashed line), and $I^{(-)}$ (triangle and dashed-dotted line) for different rotational states. The variations of $I^{(+)}$, $I^{(min)}$, and $I^{(-)}$ are shown as a function of $J$ for (a) $|M| = 0$ and $J$, as well as (b) $|M| = 1$ and $J - 1$. For $J = 16$ and $J = 10$, the three characteristic intensities are plotted against (c) even and (d) odd $|M|$ values, which correspond to even and odd ($J - |M|$) values, respectively.

**FIG. 4.** The molecular interaction potential and the resulting dipole force given by the propagating wave of $I_0 = 100 \times 10^{10}$ W/cm$^2$. Black and red curves are the variation of the negative potential and the corresponding optical dipole force, respectively, along the approximated path of the molecules occupying a specific rotational state. The potential and



the force are calculated with (solid curves) and without (dashed curves) considering the molecular alignment effect. The numbers at the top left in each graph denote the quantum numbers $J$ and $|M|$ of the rotational state. The first column contains graphs for states of different $J$ values and same $|M|$ value (namely, $|M| = 0$). The second and the third columns contain graphs for rotational states with various $|M|$ values with $J$ fixed at 10 and 20, respectively.

**FIG. 5.** The molecular interaction potential with the optical standing wave and the resulting optical dipole force. The peak intensity $I_0$ of IR1 and IR2 is set to $5.0 \times 10^{10}$, $10 \times 10^{10}$, and $20 \times 10^{10}$ W/cm$^2$ for the first, the second, and the third columns, respectively. The format of the graph is the same as that in Fig. 4, except that the potential and the force are plotted along the $x$ axis. In each column, $J$ varies from 0 to 20 while $|M| = 0$.

**FIG. 6.** The molecular interaction potential with the optical standing wave and the resulting optical dipole force for the same $J$ value (namely, $J = 10$) and different $|M|$ values (namely, $|M| = 1, 2, 9$, and 10). The format of the figure is the same as that of Fig. 5.

**FIG. 7.** Simulated velocity profiles $h(v_y)$ of the deflected molecules by the propagating laser beams with $I_0 = 10^{12}$ W/cm$^2$. The calculation was performed for the rotational temperature (a) $T = 35$ and (b) 1 K. The three profiles were calculated using the alignment-neglecting state-dependent polarizability $\alpha_{J,M}$ (black dotted curve), the alignment-including state-averaged polarizability $<\alpha^{\mathrm{F}}_{J,M}(I)>$ (blue dashed curve), and the alignment-including state-dependent polarizability $\alpha^{\mathrm{F}}_{J,M}(I)$ (red solid curve). The velocity distribution in the absence of the laser field is represented by a gray solid curve.



**FIG. 8.** Velocity $v_y$ of the $CS_2$ molecules under the influence of the propagating laser beams with $I_0 = 10^{12}$ W/cm$^2$ for (a) $J \leq 4$, (b) $J = 10$, and (c) $J = 20$. The initial conditions of the molecule – namely, $v_{0y}$, $v_{0z}$, and $y_0$ – are set to 0, $v_{mp}$, and $\omega_0/2$, respectively. The color and style of the curves representing different rotational states is the same as that in Fig. 2.

**FIG. 9.** Simulated velocity profiles $h(v_x)$ of molecules dispersed by the optical standing wave with $I_0 = 1.0 \times 10^{10}$, $5.0 \times 10^{10}$, $10 \times 10^{10}$, and $20 \times 10^{10}$ W/cm$^2$. The upper and lower panels show the velocity profiles calculated for $T = 35$ and 1 K, respectively. The coding of the curves is the same as in Fig. 7. The initial velocity distribution multiplied by 1/3 is shown by the gray solid curve in (a).




[1]     H. Stapelfeldt, H. Sakai, E. Constant, and P. B. Corkum, *Deflection of Neutral Molecules Using the Nonresonant Dipole Force*, Phys. Rev. Lett. **79**, 2787 (1997).

[2]     B. S. Zhao *et al.*, *Molecular Lens of the Nonresonant Dipole Force*, Phys. Rev. Lett. **85**, 2705 (2000).

[3]     H. S. Chung, B. S. Zhao, S. H. Lee, S. Hwang, K. Cho, S.-H. Shim, S.-M. Lim, W. K. Kang, and D. S. Chung, *Molecular Lens Applied to Benzene and Carbon Disulfide Molecular Beams*, J. Chem. Phys. **114**, 8293 (2001).

[4]     B. S. Zhao, S. H. Lee, H. S. Chung, S. Hwang, W. K. Kang, B. Friedrich, and D. S. Chung, *Separation of a Benzene and Nitric Oxide Mixture by a Molecule Prism*, J. Chem. Phys. **119**, 8905 (2003).

[5]     X. N. Sun, L. Y. Kim, B. S. Zhao, and D. S. Chung, *Rotational-State-Dependent Dispersion of Molecules by Pulsed Optical Standing Waves*, Phys. Rev. Lett. **115**, 223001 (2015).

[6]     R. Fulton, A. I. Bishop, and P. F. Barker, *Optical Stark Decelerator for Molecules*, Phys. Rev. Lett. **93**, 243004 (2004).

[7]     R. Fulton, A. I. Bishop, M. N. Shneider, and P. F. Barker, *Controlling the Motion of Cold Molecules with Deep Periodic Optical Potentials*, Nat. Phys. **2**, 465 (2006).

[8]     J. Ramirez-Serrano, K. E. Strecker, and D. W. Chandler, *Modification of the Velocity Distribution of $H_2$ Molecules in a Supersonic Beam by Intense Pulsed Optical Gradients*, Phys. Chem. Chem. Phys. **8**, 2985 (2006).

[9]     S. M. Purcell and P. F. Barker, *Tailoring the Optical Dipole Force for Molecules by Field-Induced Alignment*, Phys. Rev. Lett. **103**, 153001 (2009).

[10]    A. I. Bishop, L. Wang, and P. F. Barker, *Creating Cold Stationary Molecular Gases by Optical Stark Deceleration*, New J. Phys. **12**, 073028 (2010).





[11]    S. M. Purcell and P. F. Barker, *Controlling the Optical Dipole Force for Molecules with Field-Induced Alignment*, Phys. Rev. A **82**, 033433 (2010).

[12]    B. Friedrich and D. Herschbach, *Alignment and Trapping of Molecules in Intense Laser Fields*, Phys. Rev. Lett. **74**, 4623 (1995).

[13]    B. Friedrich and D. Herschbach, *Polarization of Molecules Induced by Intense Nonresonant Laser Fields*, J. Phys. Chem. A **99**, 15686 (1995).

[14]    H. Stapelfeldt and T. Seideman, *Colloquium : Aligning Molecules with Strong Laser Pulses*, Rev. Mod. Phys. **75**, 543 (2003).

[15]    M. Härtelt and B. Friedrich, *Directional States of Symmetric-Top Molecules Produced by Combined Static and Radiative Electric Fields*, J. Chem. Phys. **128**, 224313 (2008).

[16]    M. Lemeshko, R. V. Krems, J. M. Doyle, and S. Kais, *Manipulation of Molecules with Electromagnetic Fields*, Mol. Phys. **111**, 1648 (2013).

[17]    T. Seideman, *Manipulating External Degrees of Freedom with Intense Light: Laser Focusing and Trapping of Molecules*, J. Chem. Phys. **106**, 2881 (1997).

[18]    T. Seideman, *Shaping Molecular Beams with Intense Light*, J. Chem. Phys. **107**, 10420 (1997).

[19]    T. Seideman, *Molecular Optics in an Intense Laser Field: A Route to Nanoscale Material Design*, Phys. Rev. A **56**, R17 (1997).

[20]    E. Gershnabel and I. S. Averbukh, *Controlling Molecular Scattering by Laser-Induced Field-Free Alignment*, Phys. Rev. A **82**, 033401 (2010).

[21]    E. Gershnabel and I. S. Averbukh, *Deflection of Field-Free Aligned Molecules*, Phys. Rev. Lett. **104**, 153001 (2010).

[22]    G. Dong, W. Lu, and P. F. Barker, *Untrapped Dynamics of Molecules within an Accelerating Optical Lattice*, J. Chem. Phys. **118**, 1729 (2003).





[23]   S. Fleischer, I. S. Averbukh, and Y. Prior, *Isotope-Selective Laser Molecular Alignment*, Phys. Rev. A **74**, 041403 (2006).

[24]   B. S. Zhao, Y.-M. Koo, and D. S. Chung, *Separations Based on the Mechanical Forces of Light*, Anal. Chim. Acta **556**, 97 (2006).

[25]   S. Fleischer, I. S. Averbukh, and Y. Prior, *Selective Alignment of Molecular Spin Isomers*, Phys. Rev. Lett. **99**, 093002 (2007).

[26]   F. Filsinger, U. Erlekam, G. von Helden, J. Küpper, and G. Meijer, *Selector for Structural Isomers of Neutral Molecules*, Phys. Rev. Lett. **100**, 133003 (2008).

[27]   S. Y. T. van de Meerakker, H. L. Bethlem, and G. Meijer, *Taming Molecular Beams*, Nat. Phys. **4**, 595 (2008).

[28]   F. Filsinger, J. Küpper, G. Meijer, J. L. Hansen, J. Maurer, J. H. Nielsen, L. Holmegaard, and H. Stapelfeldt, *Pure Samples of Individual Conformers: The Separation of Stereoisomers of Complex Molecules Using Electric Fields*, Angew. Chem. Int. Ed. **48**, 6900 (2009).

[29]   F. Filsinger, J. Küpper, G. Meijer, L. Holmegaard, J. H. Nielsen, I. Nevo, J. L. Hansen, and H. Stapelfeldt, *Quantum-State Selection, Alignment, and Orientation of Large Molecules Using Static Electric and Laser Fields*, J. Chem. Phys. **131**, 064309 (2009).

[30]   S. Y. T. van de Meerakker, H. L. Bethlem, N. Vanhaecke, and G. Meijer, *Manipulation and Control of Molecular Beams*, Chem. Rev. **112**, 4828 (2012).

[31]   Y.-P. Chang, D. A. Horke, S. Trippel, and J. Küpper, *Spatially-Controlled Complex Molecules and Their Applications*, Int. Rev. Phys. Chem. **34**, 557 (2015).

[32]   M. Hillenkamp, S. Keinan, and U. Even, *Condensation Limited Cooling in Supersonic Expansions*, J. Chem. Phys. **118**, 8699 (2003).





[33]    J. Ortigoso, M. Rodríguez, M. Gupta, and B. Friedrich, *Time Evolution of Pendular States Created by the Interaction of Molecular Polarizability with a Pulsed Nonresonant Laser Field*, J. Chem. Phys. **110**, 3870 (1999).

[34]    R. Torres, R. de Nalda, and J. P. Marangos, *Dynamics of Laser-Induced Molecular Alignment in the Impulsive and Adiabatic Regimes: A Direct Comparison*, Phys. Rev. A **72**, 023420 (2005).

[35]    D. A. Horke, Y.-P. Chang, K. Długołęcki, and J. Küpper, *Separating Para and Ortho Water*, Angew. Chem. Int. Ed. **53**, 11965 (2014).

[36]    B. Friedrich and D. R. Herschbach, *Spatial Orientation of Molecules in Strong Electric Fields and Evidence for Pendular States*, Nature **353**, 412 (1991).






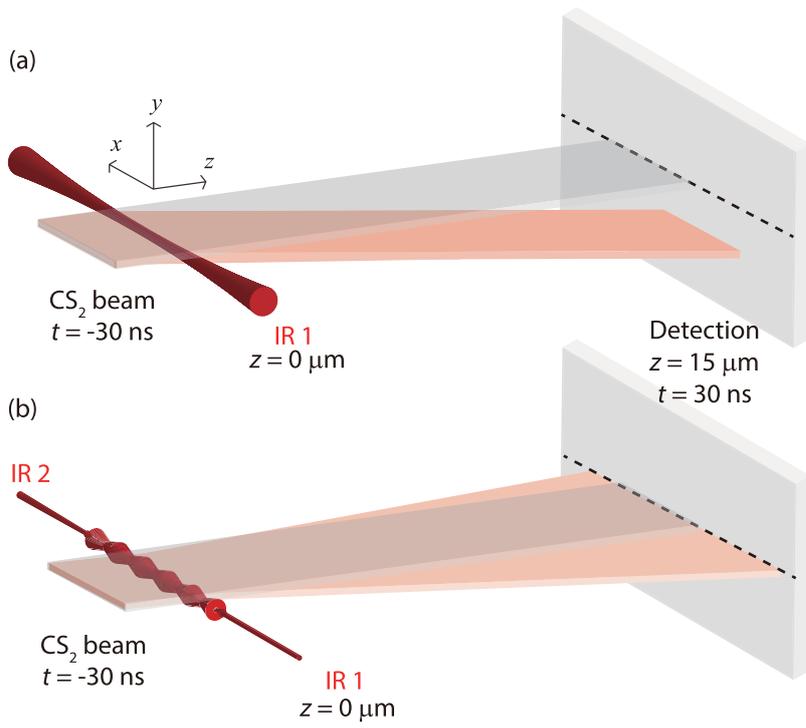

(a)

$y$
$x$ $z$

CS$_2$ beam
$t$ = -30 ns

IR 1
$z$ = 0 µm

Detection
$z$ = 15 µm
$t$ = 30 ns

(b)

IR 2

CS$_2$ beam
$t$ = -30 ns

IR 1
$z$ = 0 µm

Fig. 1



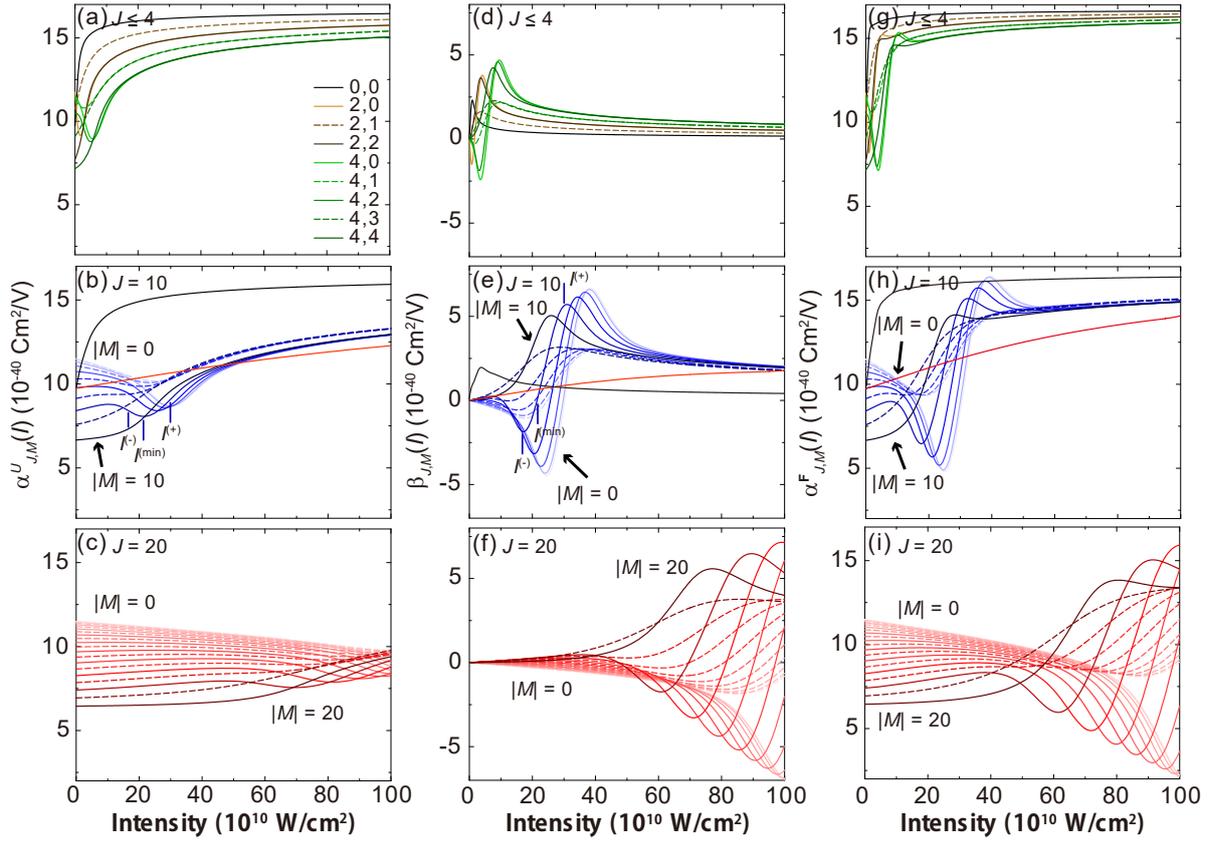

Fig. 2





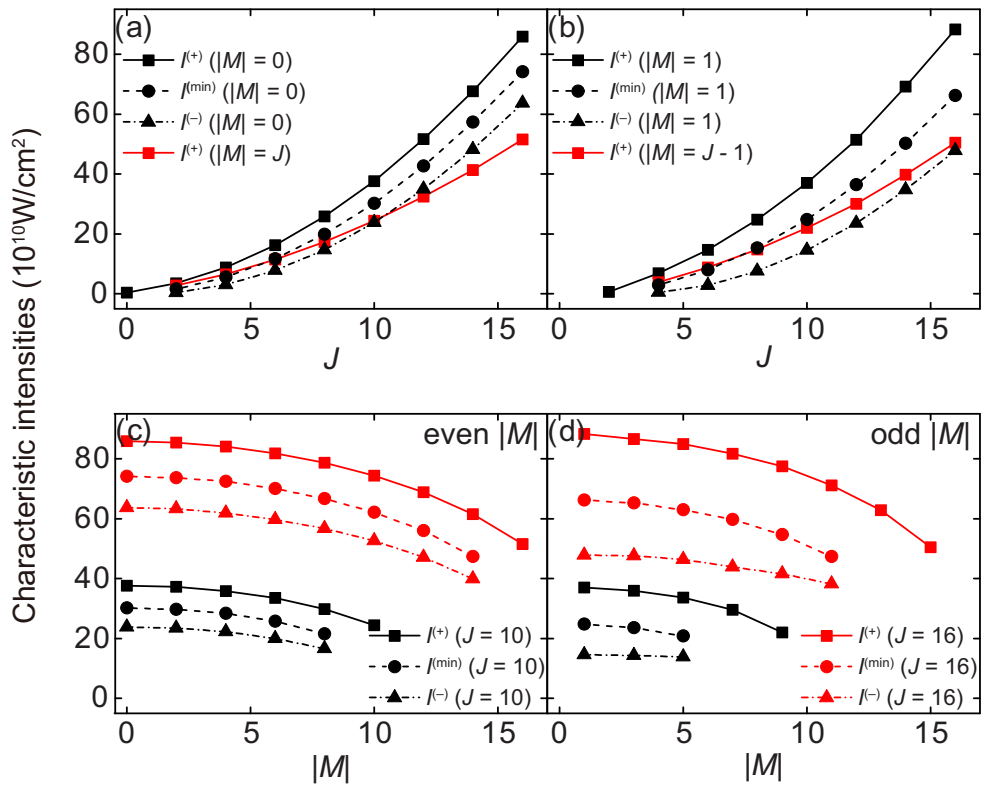

Fig. 3



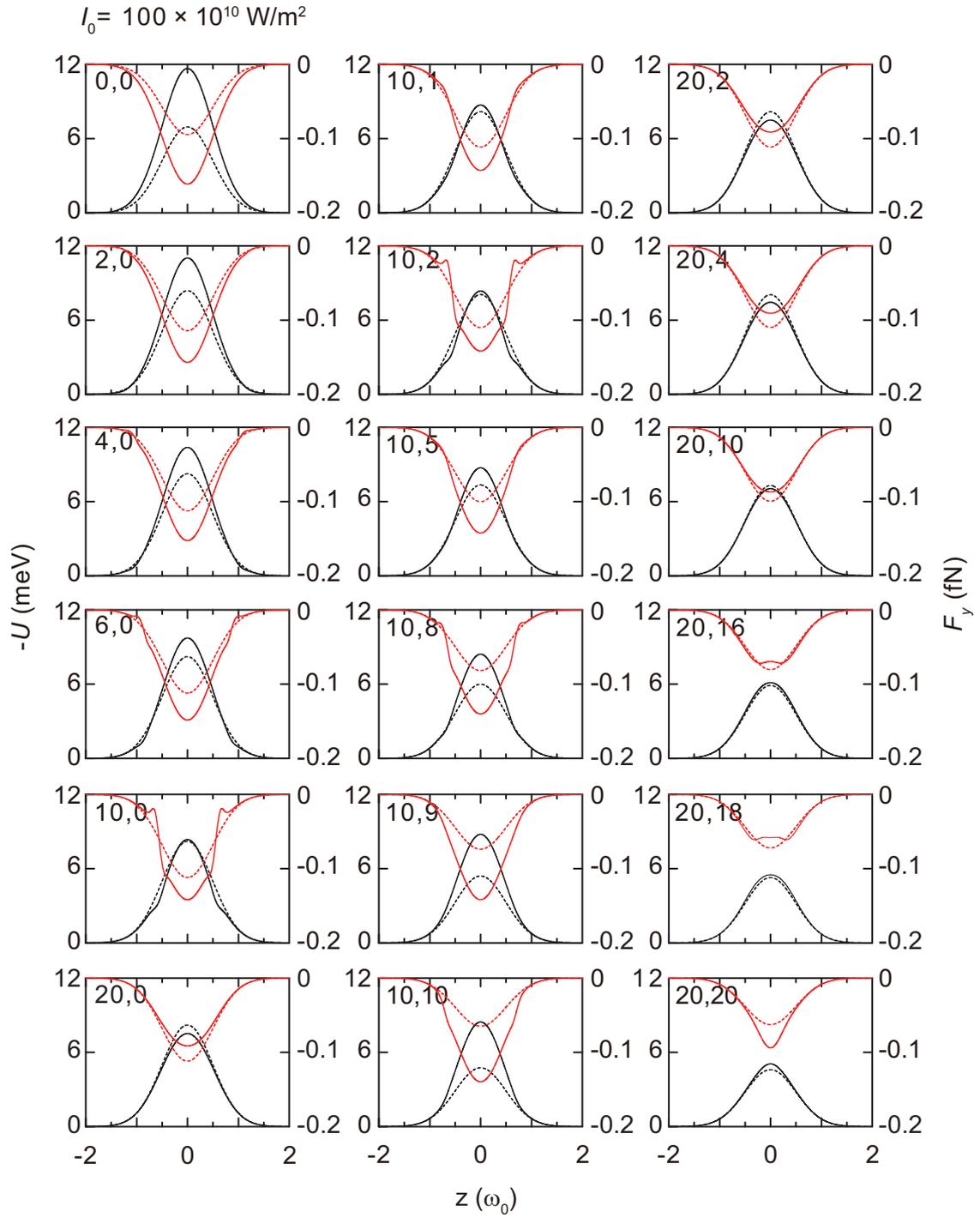

Fig. 4



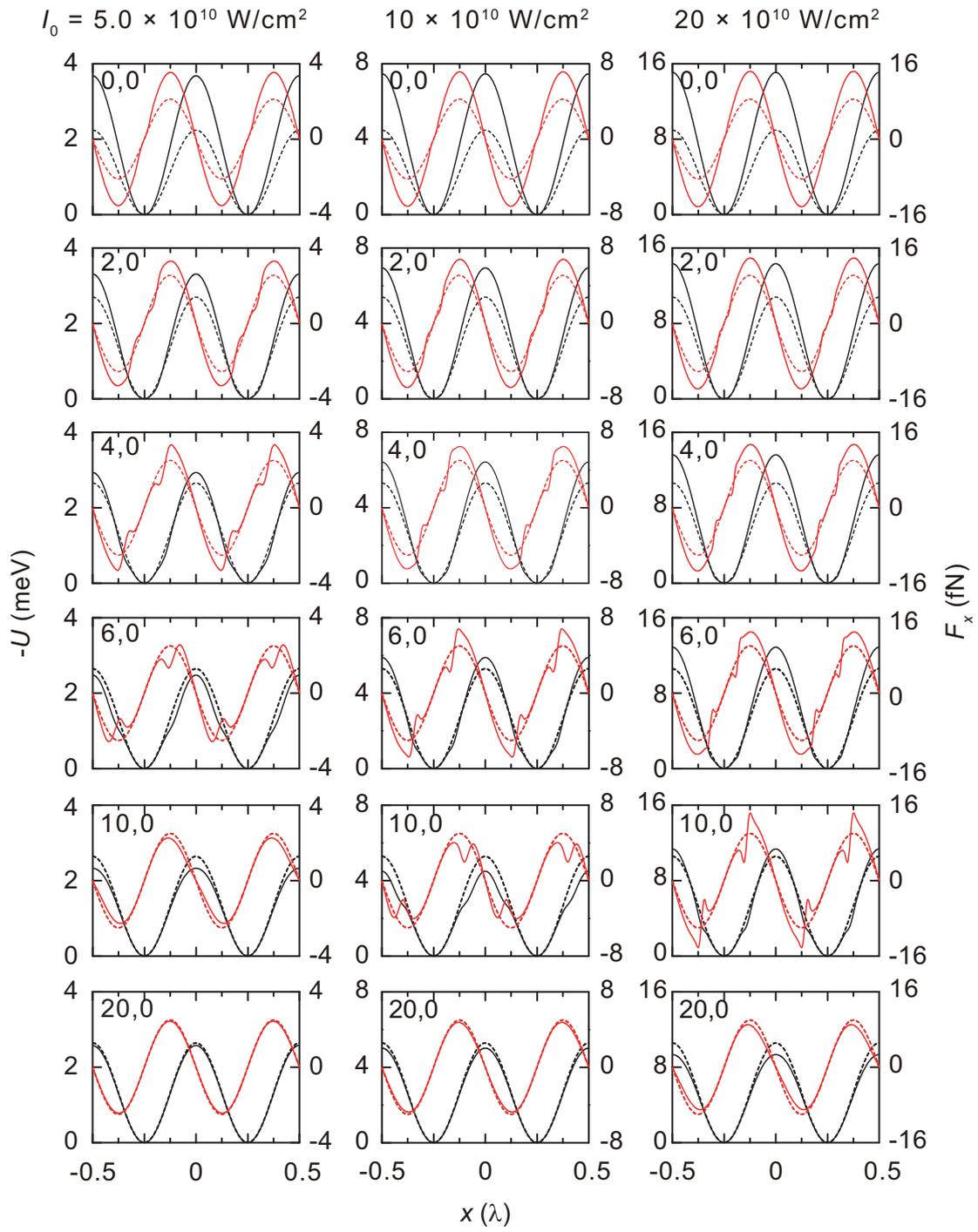

Fig. 5



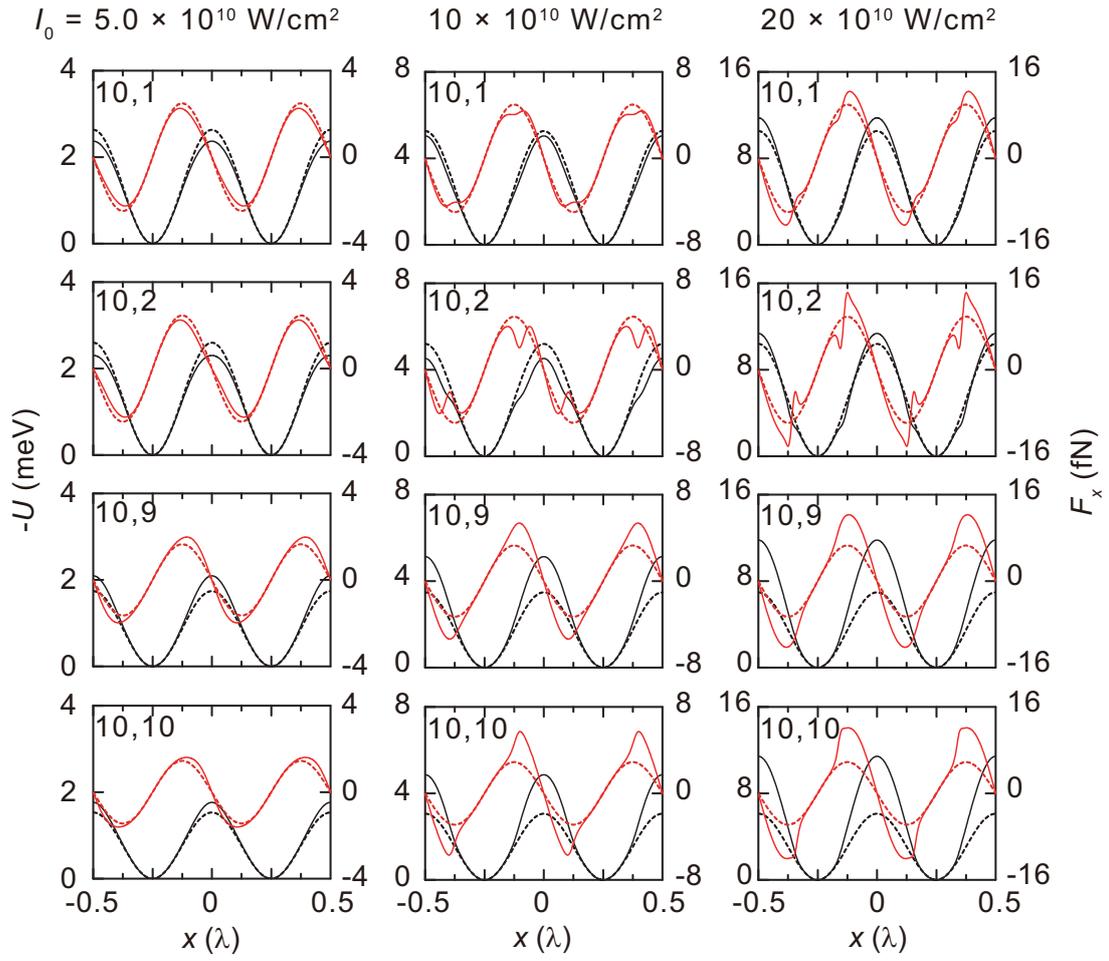

Fig. 6



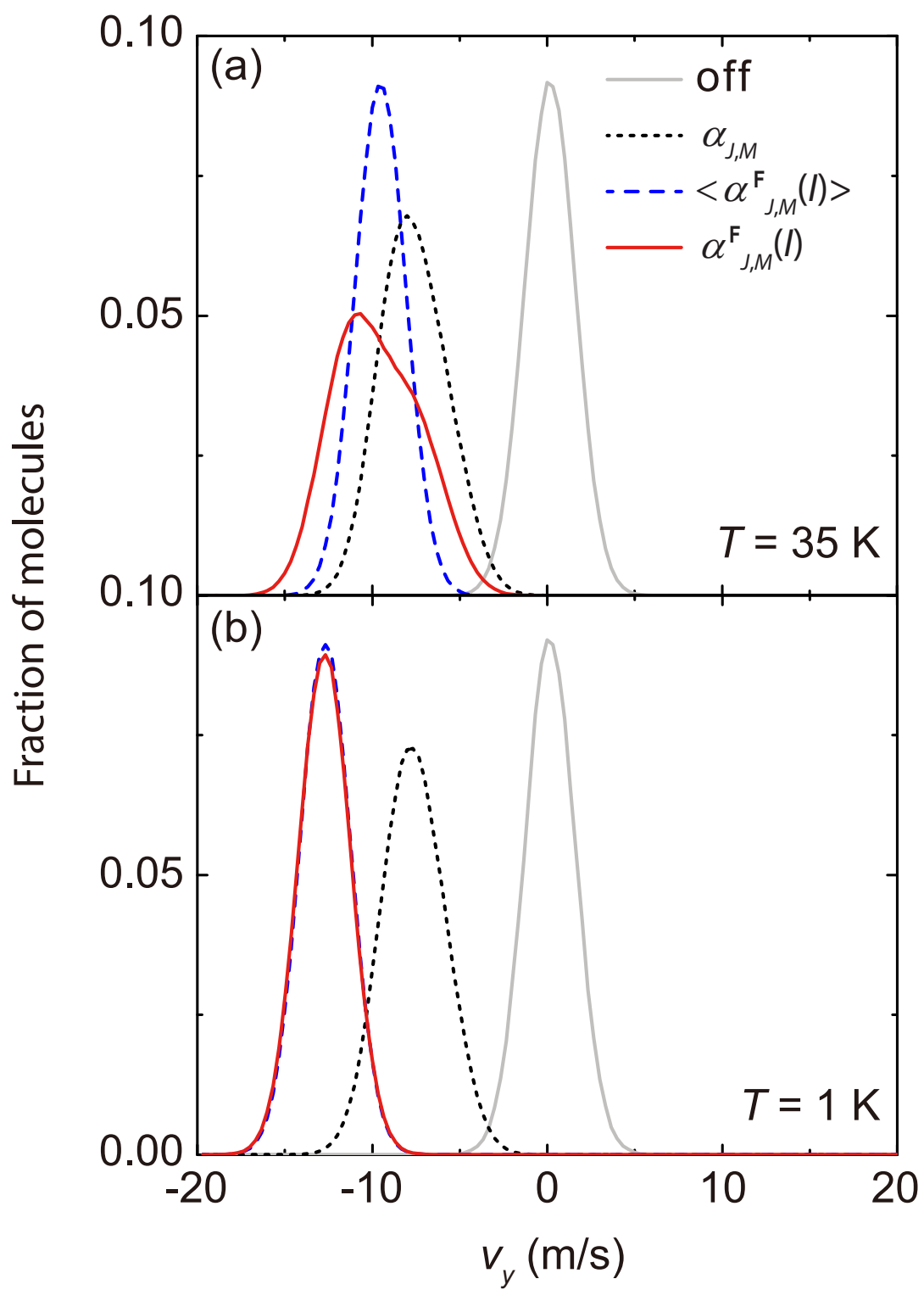

Fig. 7

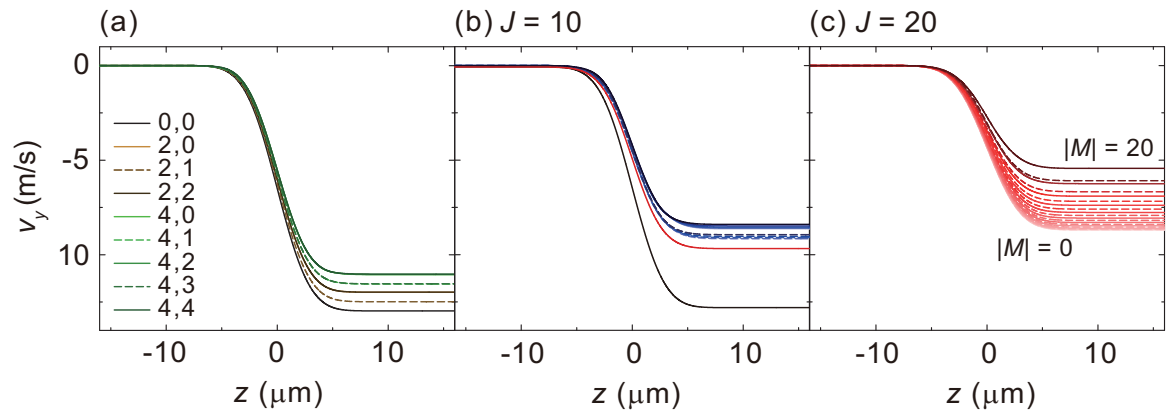

Fig. 8



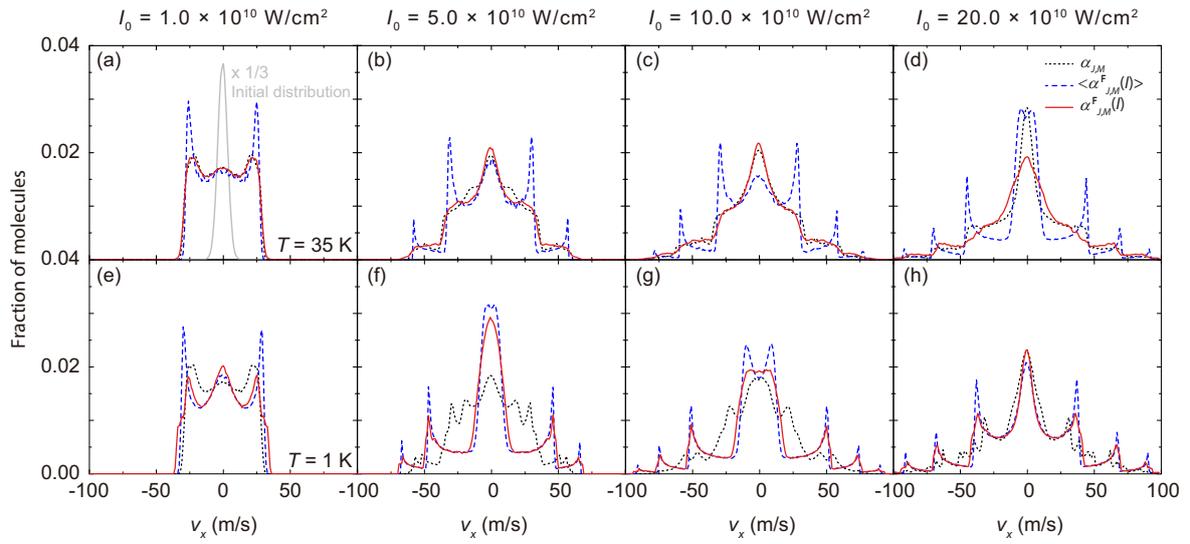

Fig. 9